\documentclass[aps,prl,twocolumn,groupedaddress,superscriptaddress,amsmath,amssymb,
]{revtex4-2}
\usepackage{xcolor}
\usepackage{graphicx}
\usepackage{dcolumn}
\usepackage{bm}

\begin{document}

\title{Buckling instability in a chain of sticky bubbles}

\author{Carmen L. Lee}
\email{Current address: Physics Department, North Carolina State University, Raleigh, NC, USA}
\affiliation{Department of Physics \& Astronomy, McMaster University, Hamilton, Ontario, L8S 4M1, Canada}

\author{Kari Dalnoki-Veress}
 \email{Corresponding author: dalnoki@mcmaster.ca}
\affiliation{Department of Physics \& Astronomy, McMaster University, Hamilton, Ontario, L8S 4M1, Canada}
\affiliation{Gulliver, CNRS UMR 7083, ESPCI Paris, Univ. PSL, 75005 Paris, France.}

\date{\today}%

\begin{abstract}
A slender object undergoing an axial compression will buckle to alleviate the stress. Typically the morphology of the deformed object depends on the bending stiffness for solids, or the viscoelastic properties for liquid threads. We study a chain of uniform sticky air bubbles that rise due to buoyancy through an aqueous bath. A buckling instability of the bubble chain with a characteristic wavelength is observed. If a chain of bubbles is produced faster than it is able to rise, the dominance of viscous drag over buoyancy results in a compressive stress that is alleviated by buckling the bubble chain. Using low Reynolds number hydrodynamics, we predict the critical buckling speed, the terminal speed of a buckled chain, and the geometry of the buckles.
\end{abstract}

\maketitle

\begin{figure}
\includegraphics[width = 1\columnwidth]{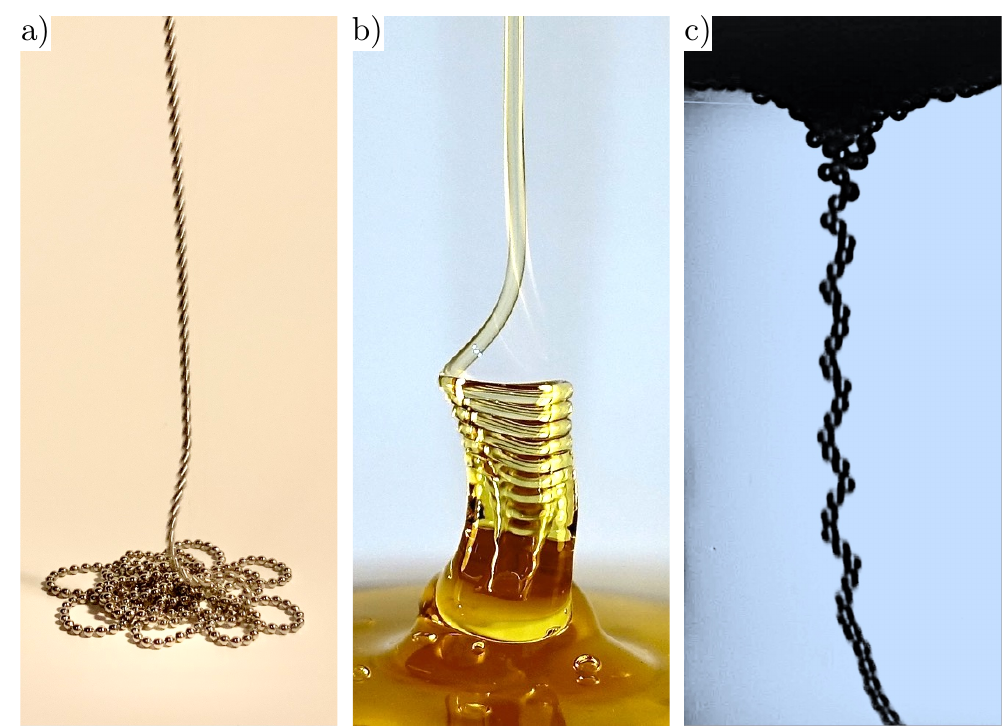}
\caption{\label{f:coils} Buckling instabilities resulting in coiling: (a) Buckling of bead chain dropped at $\sim 0.2$~m/s (bead diameter 1.5~mm). (b) Viscous coiling of Lyle's golden syrup (liquid thread diameter $\sim 1$~mm). (c) Image from the experiment showing the buckling of a chain of air bubbles in an aqueous fluid rising due to buoyancy and collecting at the air/bath interface at the top (bubble diameter $\sim 50 \ \mu$m).}
\end{figure}

Slender objects, like strands of hair, rope or blades of grass easily buckle when compressed along the axial direction. Buckling can occur on a multitude of length scales from macroscopic, like a rope coiling when hitting the ground~\cite{Mahadevan1996, Jawed2014} as shown in Fig.~\ref{f:coils}(a) for a falling chain; to microscopic, like the bending of flagella while micro-organisms swim~\cite{Gray1955, Chwang1971,duRoure2019, Purcell1997}. If the stress on the slender object is normal to the cross-section of the object, the object will undergo a regular deformation, for example a rope fed at a constant speed into a cylindrical tube will bend with a characteristic wavelength determined by the friction and bending stiffness of the rope~\cite{Miller2015}. Interestingly, this phenomenon is not limited to solid materials and can also be seen with viscous  jets. Take for instance the familiar example of a stream of honey which coils when it falls onto  toast shown in Fig.~\ref{f:coils}(b)~\cite{Skorobogatiy2000, Ribe2006, Barnes1958, Chakrabarti2021,Habibi2014, RibeJFM2006, Mahadevan1998, Ribe2004}. Here the viscosity of the liquid resists the bending of the thread, and a regular coiling is observed~\cite{Mahadevan1998}. In addition to buckling and coiling due to the compressive stress induced by a barrier, viscous drag can induce buckling of slender structures driven through a viscous liquid, as shown in the work by Gosselin \textit{et al.} for solid threads~\cite{Gosselin2014}, and Chakrabarti \textit{et al.} for gelling structures~\cite{Chakrabarti2021}. Coiling and buckling arises in diverse areas from orogeny in geosciences, to the coiling of DNA structures,  and is of common concern to those building architectural structures~\cite{Ribe2012}. 
Furthermore, coiling and buckling of slender fibers has continued to be explored for applications in 3D printing, preparation of metamaterials, and electrospinning on scales ranging from centimetric to nanometric~\cite{Mehdipour2020, Nunes2013,Kim2010, Brun2015, Jawed2015,Liu2018, Han2007}. 
 
\begin{figure*}
\includegraphics[width =1\textwidth]{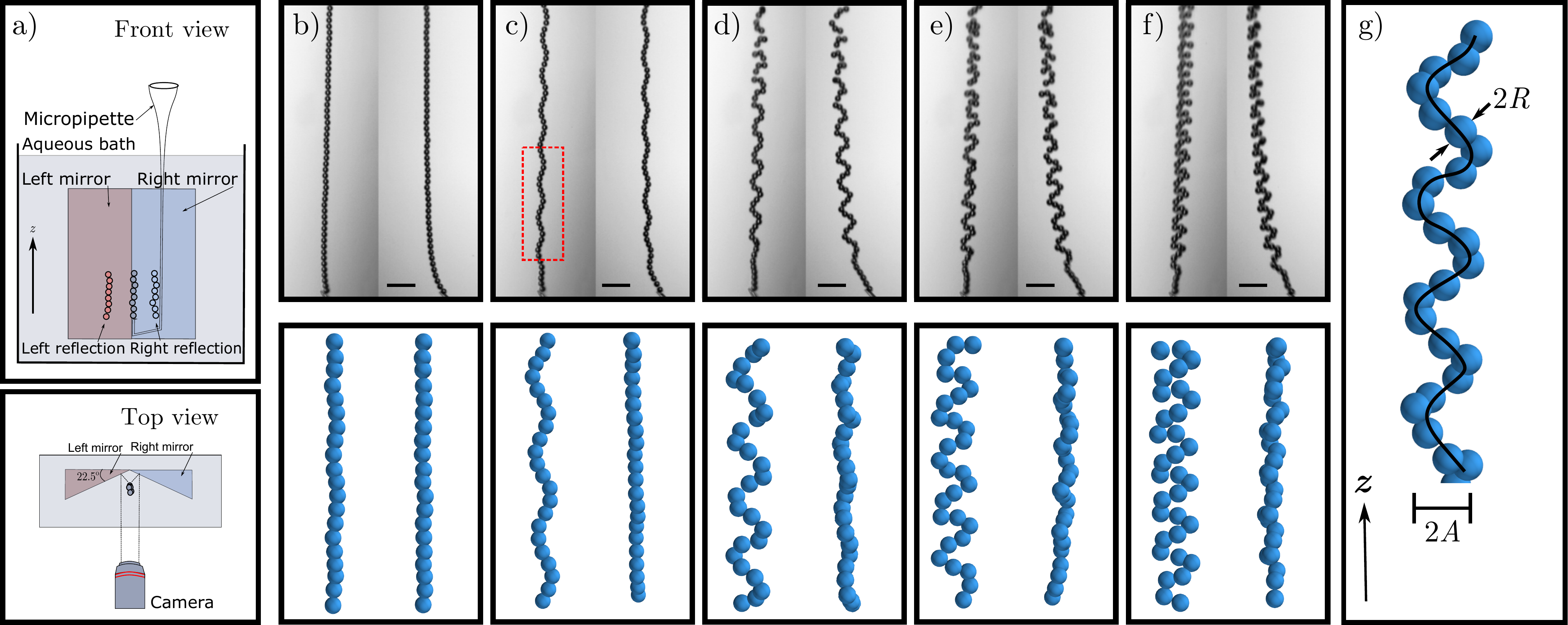}
\caption{\label{f:images} (a) Schematic of the bubble chamber and imaging set-up. Front view shows the micropipette, and a short section of bubbles to illustrate the reflections. Top view shows the mirror angles used to image the two orthogonal planes. (b)-(f) Images of a bubble chain undergoing a buckling instability as the bubble production speed is increased from left to right ranging from 7.8 mm/s to 27.2 mm/s. Each panel shows images of the same chain taken at two orthogonal planes to allow for 3-dimensional reconstruction. Faster bubble production speed increases the buckling amplitude. The scale bar corresponds to 250 $\mu$m. The 3-dimensional reconstruction of a short segment (size of reconstruction is shown in panel (c) by the red dashed box) of the bubbles is shown below the corresponding experimental images (b)-(f). Reconstructions are rotated to show the angle with maximal buckling amplitude (left side), and the corresponding orthogonal angle. The schematic shown in (g) provides the relevant geometric values: bubble radius $R$, amplitude $A$, and coordinate axis $z$.}
\end{figure*}

In addition to the two examples shown in  Fig.~\ref{f:coils}(a) and (b), in panel (c) we see the buckling of a chain of air bubbles that rise due to buoyancy, which is the subject of this study. Previous studies have focussed on the shape~\cite{tripathi2015dynamics, lalanne2020determination},  and trajectories~\cite{mougin2001path, mathai2018flutter, shew2006dynamical,Atasi2023} of rising bubbles and jets of bubbles, as well as the role of droplets and bubbles in the dynamics of multiphase systems \cite{verschoof2016bubble,Guzowski2022,Feneuil2023}. Recent work by Atasi \textit{et al.} is an example of the complexities encountered with jets of bubbles and the role of surfactants in the liquid phase. Here we experimentally investigate the buckling of a chain of adhesive, uniform air bubbles in an aqueous bath. The sticky bubbles are produced at the bottom of the chamber from a small orifice [see schematic in Fig.~\ref{f:images}(a) and images (b)-(f)]. The adhesion between the bubbles is crucial: by producing the bubbles quickly such that each subsequent bubble is produced before the previous bubble has risen by a distance of one diameter, the two adhere due to short-range depletion forces. Producing multiple bubbles in a row creates a linear chain [Fig.~\ref{f:images}(b)].  Upon increasing the bubble production speed further, the hydrodynamic drag force increases; and, at some point exceeds the buoyant force for a given length. At this point the linear chain is no longer stable, and buckling is induced by a compressive force due to viscous drag acting on this granular system of sticky bubbles [Fig.~\ref{f:images}(c)]. 

Previous work on the buckling of solid and liquid threads~\cite{Mahadevan1996, Mahadevan1998, Chakrabarti2021}, found that a bending resistance is critical to the phenomenon. However, with the bubble chain there is no intrinsic cost to bending: the bubbles have no solid-solid friction, nor do the bubbles have the viscous resistance of a liquid thread to bending; yet, a characteristic buckling length emerges. The physical mechanism for the buckling seen in Fig.~\ref{f:coils}(a) and (b) is fundamentally different to that observed in the bubble chain. We first investigate the balance between the hydrodynamic drag and buoyancy to obtain a criterion for the onset of buckling. We then explore the relationship between bubble size, production speed, and viscosity, and determine the terminal velocity of the rising undulated chain of bubbles. Finally, we use hydrodynamic drag to explain the dependence of the buckling amplitude and wavelength on the rate at which the bubble chain is produced. The model, which relies on simple geometric and hydrodynamic arguments, is sufficient to explain the experimental results.

Typical images of the buckling experiment are shown in Fig.~\ref{f:images}(b)-(f) (see video in the Supplemental Material~\cite{sm}), with relevant parameters shown in Fig.~\ref{f:images}(g). 
A chain of bubbles with radius $R$ is produced at speed $q$, and rises with terminal speed $v$, and a buckling amplitude $A$.  The air bubbles are prepared by pushing air though a small glass micropipette with an opening (diameter $\sim$ 10 $\mu$m) into an aqueous bath with surfactant, sodium dodecyl sulfate (SDS) and salt (NaCl) (see Appendix A). The SDS concentrations used in these experiments range from 0.035 M to 0.28 M, and are well above the critical micelle concentration (CMC)~\cite{Woolfrey1986}. We emphasize that the surfactant serves two purposes in our experiment. First, the surfactant stabilizes the bubbles against coalescence. Second, excess SDS forms micelles in the solutions which controls the adhesion between the bubbles via the depletion interaction~\cite{OnoditBiot2020}.  The adhesion stabilizes the chain against breaking apart due to viscous stresses and buoyancy. The differing SDS concentrations are accompanied by a small change in the viscosity of the aqueous bath~\cite{Poskanzer1975} due to the presence of micelles~\cite{RuizMorales2018}. The viscosity of the three solutions was measured independently and are 1.5, 1.6 and 2.0~mPa$\cdot$s (see Appendix A). The solution has a density $\rho \approx 1 \mathrm{\,kg/m}^3$.  NaCl is added to the solution to screen electrostatic interactions.  The pressure through the micropipette is kept constant for the trials by using a syringe and plunger. For the small amounts of air being expelled through the micropipette relative to the air volume in the syringe, we can treat the amount of air as an infinite reservoir with a constant pressure, which results in a constant bubble size~\cite{Tate1864, Yuan2016}. Changing the size of the micropipette orifice creates bubbles with radii ranging from $16~\mu\textrm{m} < R < 38~\mu$m. At these small length scales, the bubbles have a large enough Laplace pressure to be treated as hard spheres. In these experiments, the Reynolds number is Re = $\rho R v/\mu <$ 1 for all 105 experiments, and viscous forces dominate over inertial ones. 

The aqueous bath is contained in a cell with two mirrors set at 22.5$^\circ$ with respect to the back plane of the reservoir so that one camera can simultaneously image two orthogonal planes of the bubble chain [see Fig.~\ref{f:images}(a)].  The orthogonal views (Fig.\ref{f:images}(b)-(f) are used to reconstruct the three-dimensional shape of the chain (Appendix B). The reconstructions are shown in Fig~\ref{f:images} below each corresponding experiment in (b)-(f); however, the reconstructions are rotated so that the maximal amplitude of the chain is shown on the left, and the minimal amplitude on the right. It is clear that the buckling takes place predominantly in a two dimensional plane. We do not observe stable helix formation and suspect that this is due to the small compression accessible in the experiment~\cite{Chakrabarti2020} or hydrodynamic stabilization (entrainment) of the two-dimensional structure. There is a symmetry-breaking which sets the buckling plane, and from our experiments we find that this is determined by the angle at which the bubbles emerge from the orifice, or any small oscillations in the pipette.

\emph{Critical production speed for buckling onset} -- We first focus on the critical speed at which the bubble chain buckles. At low bubble production speeds $q$, the adhesive bubbles naturally align in the vertical direction due to the buoyant force $\vec{F}_b$. In fact, if stationary, the chain is under tension due to buoyancy. However, since the bubble chain is created with a speed $q$, there is also a drag force acting downwards $\vec{F}_d(q)$ which depends on $q$. The tension in the chain switches to a compressive force at a critical speed $q_c$ when the magnitude of the drag force of the chain exceeds the magnitude of the buoyant force. With no bending stiffness to the chain, a compressive force acting on the chain is the minimal requirement for the chain to buckle and therefore the chain will buckle at critical speed $q_c$, where the chain transitions from being in tension to being in compression, or when $\vec{F}_b+\vec{F}_d(q)=0$.

\begin{figure}[]
\includegraphics[width = 0.48\textwidth]{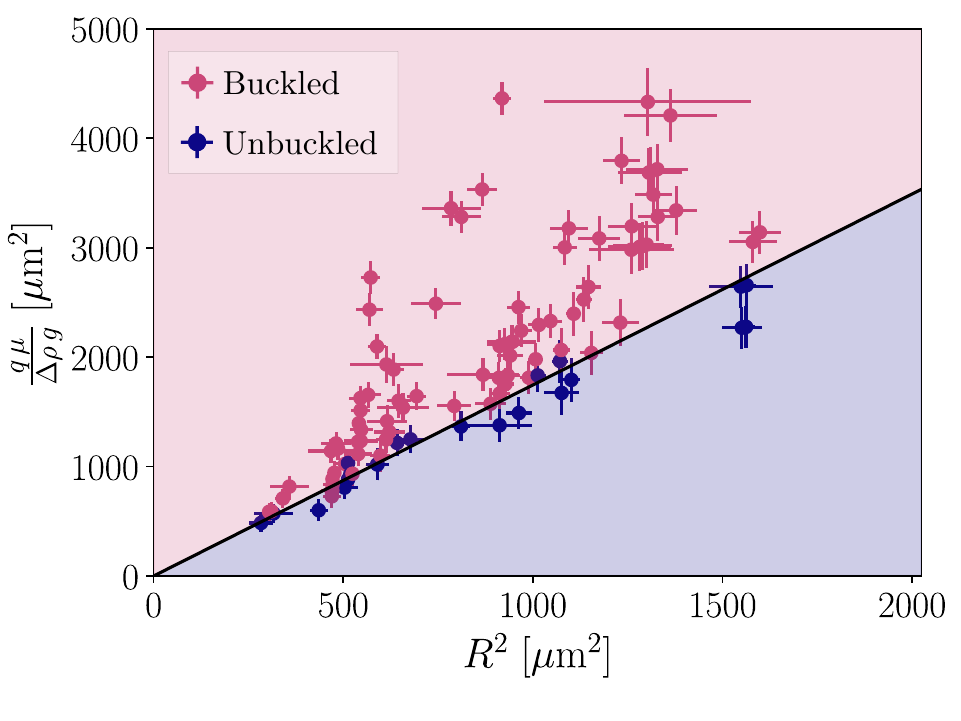}
\caption{\label{f:phase} A phase diagram showing the chain production speed $q$ normalized by the ratio of inertial forces to viscous forces, as a function of the bubble radius $R$ squared. The dark blue data points indicate an unbuckled chain, where the light pink data points indicate buckled chains. The error bars indicate the measurement uncertainty in the radius of the bubbles, the viscosity and chain production speed. The black line corresponds to the predicted theoretical relationship in Eq.~\ref{eq:critq}.}
\end{figure}

For a given section of chain with $n$ bubbles and length $l=2Rn$, the force due to buoyancy is $\vec{F}_b = 4\pi R^3\Delta \rho g n/3$; where $\Delta \rho$ is the difference in density between the air and the bath and $g$ is the acceleration due to gravity. Given the small Reynolds number in these experiments, the hydrodynamic drag takes the general expression of $\vec{F}_d = - c_c q  \mu  l$; where $c_c$ is a dimensionless drag coefficient tangential to the chain which is of order 1 using slender body resistive force theory~\cite{Huner1977, Ui1984, Gray1955, Lighthill1976}. Thus our criterion for buckling is given by $4\pi R^3\Delta \rho g n/3- c_c q_c  \mu  l=0$,  and we obtain:
\begin{equation}
    q_c = \frac{2 \pi  R^2  \Delta\rho  g}{3 c_c  \mu}.
    \label{eq:critq}
\end{equation}
Following Eq.~\ref{eq:critq}, in Fig.~\ref{f:phase} we plot $q \mu/ \Delta\rho g$ as a function of $R^2$. Unbuckled (dark blue) and buckled (pink) chains are plotted to form a phase diagram bounded by the line given by Eq.~\ref{eq:critq}. As expected, for  $q<q_c$  the chain is in tension and no buckling is observed, while for $q>q_c$ buckling is observed. Since the slope of the phase boundary is given by $2 \pi/3 c_c$, the only fit parameter is the dimensionless drag coefficient $c_c= 1.2 \pm 0.1$, which compares well to the expectation of a constant of order 1~\cite{Huner1977, Gray1955, Lighthill1976}. The excellent agreement between the data and theory indicate that we can predict the critical bubble production speed for the onset of buckling in a chain of bubbles. We note again that here there is no bending resistance and the buckling onset is the result of hydrodynamics. 

\emph{The terminal velocity of the chain} -- As the bubbles are produced, the effect of the increased production speed is only sustained over the first small number of bubbles due to hydrodynamic drag; this is sensible and consistent with the observation that when one tries to push the end of a rope, the buckling happens near the location of the push. Well above the bubble production orifice, each of the bubbles only moves in the vertical direction and does not translate along the arc length of the bubble chain \textit{i.e.} each bubble only has a vertical component to its velocity. Since the Reynolds number is small for this system and there will be fluid entrained between the spaces of the bubbles, we approximate the buckled chain as a slender ribbon moving upwards through the liquid. Because the chain is moving with a constant speed, we  balance the  buoyant force acting on the bubbles with the drag force acting on the ribbon to obtain a terminal velocity $v$.

We consider the buoyant force per unit length along the vertical direction of the chain and write this as:
$
\frac{\vec{F}_b}{z}=\frac{4 \pi R^3\Delta \rho g}{3} \cdot \frac{dn}{dz};
$
where $dn/dz$ represents the number of bubbles per unit vertical length. Since the bubble chain is being produced with a speed $q$ and the ribbon of chain is moving vertically with a speed of $v$, we have that:
$
\frac{dn}{dz}=\frac{q}{v}\cdot \frac{1}{2R}.
$
We can approximate the drag on the ribbon, using resistive force theory as:
$
\frac{\vec{F}_d}{z}=-c_r\, \mu\, v;
$
where $c_r$ is the drag coefficient of the ribbon. Again, we expect the drag coefficient to be of order 1; however, it is a strong assumption to take $c_r$ constant irrespective of the aspect ratio of the ribbon. Nevertheless, the corrections are typically small~\cite{Borker2019}, especially since here the maximum aspect ratio of the ribbon cross-section is $\sim4$ (maximum amplitude is $\sim$ 2 bubbles). We find that the assumption is sufficient to capture the essential physics as will be clear below. Setting $\vec{F}_b+\vec{F}_d=0$ we obtain the terminal velocity:
\begin{equation}
    v=\left(\frac{2\pi}{3c_r}\cdot \frac{R^2\, q\, \Delta \rho\, g}{\mu}\right)^{1/2}.
\label{eq:speed}
\end{equation}

\begin{figure}[]
\includegraphics[width = 1\columnwidth]{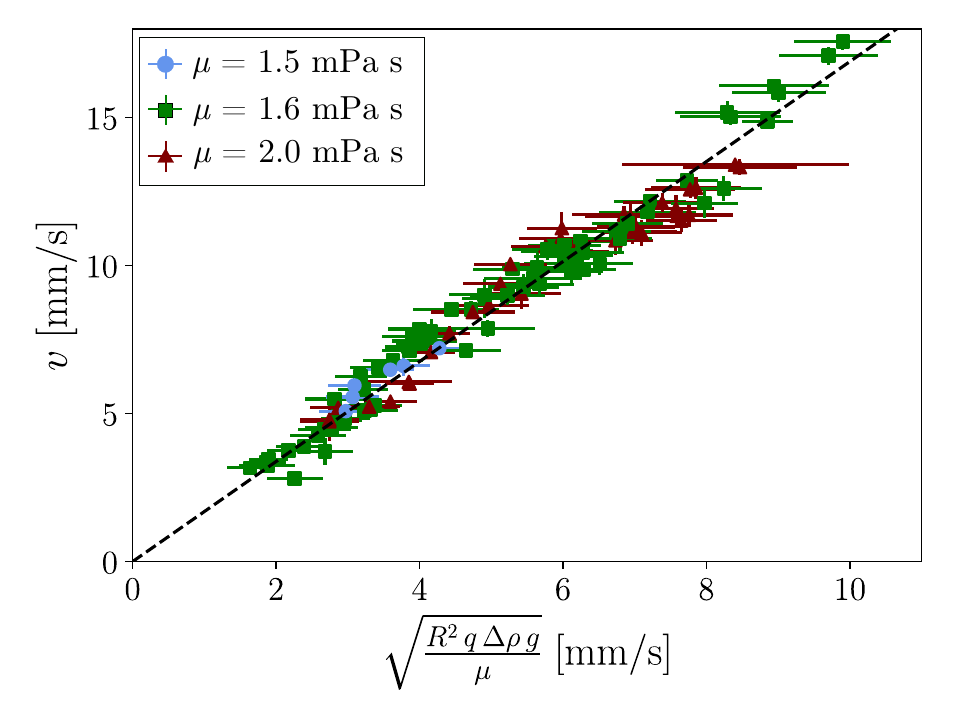}
\caption{\label{f:speed}  
Plot of  $v$ as a function of $\left(R^2 q \Delta \rho g /\mu\right)^{1/2}$ in accordance with Eq.~\ref{eq:speed} to test the dependence of the terminal speed of the undulated chain of bubbles on the experimental parameters. Bubble sizes range from $\sim 16 \ \mu\textrm{m} < R < 38 \ \mu$m, viscosities are indicated in the legend and correspond to SDS concentrations of 0.035 M, 0.14 M and 0.28 M. The black dashed line is the best fit of Eq.~\ref{eq:speed}.
}
\end{figure}

To test Eq.~\ref{eq:speed}, we plot the velocity $v$ as a function of $\left(R^2 q \Delta \rho g /\mu\right)^{1/2}$, for various values of $q$, $R$ and $\mu$. Although the change in viscosity is relatively small (1.5, 1.6 and 2.0~mPa$\cdot$s), the drag depends on viscosity and must be taken into account (see Eq.~\ref{eq:speed}). As expected, we see in Figure \ref{f:speed} that the terminal speed increases with increasing chain production speed. 
The expected relationship defined in Eq.~\ref{eq:speed} and the measured data show excellent agreement, with the only free parameter being the drag coefficient for the ribbon $c_r$ = 0.72 $\pm$ 0.09. Although the hydrodynamic drag has been modelled with a single drag coefficient $c_r$, clearly the assumption proves to be an adequate approximation for the undulating chain with small amplitudes shown in Fig.~\ref{f:images}, and we find that $c_r \approx c_c$. 

\emph{Buckling amplitude and wavelength} -- From Fig.~\ref{f:images}, one can see that the buckling amplitude $A$ increases with increasing chain productions speed $q$, while the wavelength, $\lambda$, decreases with increasing $q$. The buckling results from an excess of chain length that is produced relative to the speed that the chain can move upwards. We make the simple \emph{ansatz} that the amplitude must scale as $q-q_c$ since buckling can only happen when $q>q_c$ (i.e. $A=0$ for $q<q_c$).  Because $q-q_c$ is a speed, we must multiply by a characteristic timescale, $\tau$, to get a lengthscale. Given that the relevant parameters in the problem are $R$, $\Delta \rho$, $g$, and $\mu$, we have $\tau \sim \frac{\mu}{R \Delta \rho g}$ and can write the amplitude as~\cite{tau}: 
\begin{equation}
   A \propto (q-q_c) \cdot \frac{\mu}{R \Delta \rho g}.
   \label{eq:amplitude}
\end{equation}

\begin{figure}
\includegraphics[width = 0.48\textwidth]{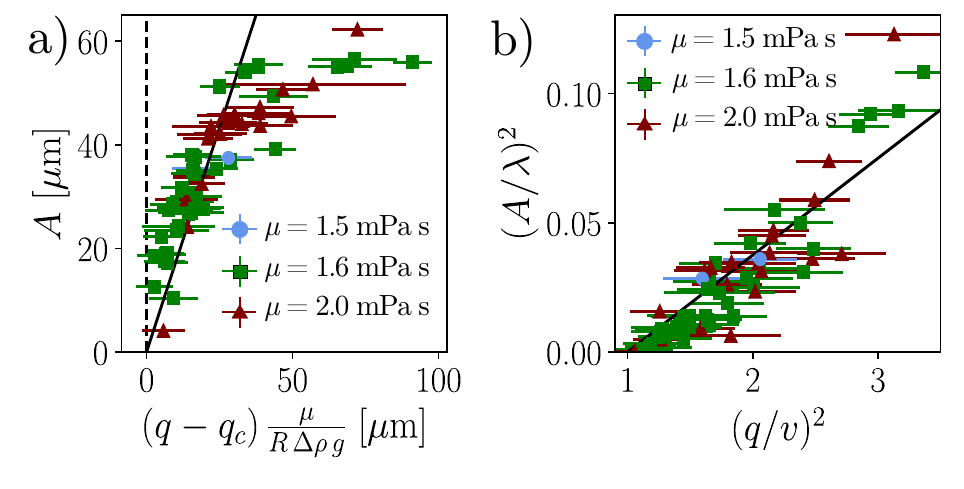}
\caption{\label{f:amplitude} 
a) The buckling amplitude $A$ as a function of $(q-q_c) \mu/R \Delta \rho g$ collapses the data in accordance with Eq.~\ref{eq:amplitude}. The solid black line shows the line of best fit to the small amplitude data with a slope of $1.8\pm 0.6$. b) The relation between the ratio $A/\lambda$ and $q/v$. The black line shows the model prediction with a slope of $0.04 \pm 0.01$. Bubble sizes range from $16 \ \mu\textrm{m} < R < 38 \ \mu$m, viscosities are indicated in the legend. 
}
\end{figure}

In Fig.~\ref{f:amplitude}a) we test Eq.~\ref{eq:amplitude} and plot $A$ as a function of $(q-q_c) \mu/R \Delta \rho g$. For a variety of bubble sizes, bubble production speeds, and viscosities, the experimentally obtained amplitude closely agrees with the predicted relationship for all but the largest amplitudes where the simple scaling model fails. At large differences between the production speeds we see a deviation from the prediction, indicating that the simple scaling is insufficient to capture the unaccounted non-linear behaviour at the largest production speeds. 

The shape of the buckled chain can equivalently be described by the wavelength using simple geometric relationships between $A$, $\lambda$, $v$, and $q$. In order to obtain an analytic solution, we approximate the undulatory ribbon as a sawtooth profile. The approximation introduces an error that is at most $\sim 3.5$\%, in comparison to a more rigorous arc-length of a sinusoid, but yields an analytic solution: 
We relate the time that the chain rises by a distance of $\lambda$, to the arc-length of chain produced in that time. Then the arc-lenght of chain, per unit wavelength is $l_\lambda/\lambda=q/v$. For a sawtooth profile, we obtain the simple expression $16 (A/\lambda)^2=(q/v)^2-1$. We note that the ratio $A/\lambda$ vanishes when $q=v$, this is intuitive since $A \rightarrow 0$ and $\lambda \rightarrow \infty$ as the bubble production speed decreases to the velocity at which the chain rises: the chain becomes straight. In Figure~\ref{f:amplitude}b) we see that this simple relationship is in good agreement with the data with a best fit slope of $~1/27$  which differs from simple geometric model by $\sim 1.6$. We attribute the difference due to the sawtooth approximation, as well as differences in the measurement  of $A$ and $\lambda$ (see Appendix B).

\emph{Conclusions:} We have studied the buckling instability that a slender chain of bubbles undergoes when travelling through a viscous bath. The instability arises when the drag force of the rising chain of bubbles exceeds the buoyant forces.  After the buckles are formed, the chain moves as a ribbon through the bath and the terminal velocity can be calculated using a simple balance between buoyancy and drag. We find that the model well describes the experiments for the buckling onset, predicted terminal velocities, the amplitude of the buckling, and the relationship between the amplitude and the wavelength. We have studied a system that shows buckling that is akin to many other familiar systems, like honey coiling on toast, the coiling of a falling rope, or the bending of a fiber pushed into a viscous fluid. However, the fundamental origin of the buckling is fundamentally different. Unlike other systems, in which buckling arises from a cost associated with bending, to our knowledge this is the first study of drag-induced buckling with no intrinsic cost to bending -- a buckling instability with a characteristic length-scale emerges as a result of hydrodynamics.

\acknowledgements{C.L.L. and K.D.-V. acknowledge financial support from the Natural Science and Engineering Research Council of Canada. C.L.L. acknowledges funding from the Vanier Canada Graduate Scholarship.}

\section{Appendix A: Bubble production}
 The bubbles are created by forcing air through a pulled glass micropipette into an aqueous bath composed of water (HPLC, Sigma-Aldrich), NaCl (Caledon, 1.5\% w/w), and sodium dodecyl sulfate (SDS, Bioshop, with concentration ranging from 0.035 M to 0.28 M or 1\% to 8\% w/w). We note that at these high SDS concentrations, the surface tension is  independent of the concentration and can be assumed constant~\cite{Woolfrey1986}. To create small air bubbles such that the dynamics are slow enough for the Reynolds number to be small, the orifice to produce the bubbles must also be small \cite{Yuan2016, Tate1864}.  Glass capillary tubes of initial thickness 1 mm in diameter (World Precision Instruments Inc.) are heated and pulled with a pipette puller (Narishige PN-30). After pulling, the tip of the pipette is long and flexible and has a diameter of ~30 $\mu$m. The air bubbles produced by these pipettes are still too large. To produce smaller bubbles, the pipettes are manually pulled a second time. Here, we heat the tip of the pipette by putting it into contact with a hot platinum wire where the glass locally melts and sticks to the wire. The pipette is pulled from the wire using the natural flexibility of the pipette, forming a narrower cone at the end with diameters ranging from  $\sim$ 10 $\mu$m to 25 $\mu$m..

The viscosity of the aqueous bath is calculated using Stokes' law for the rise of a single bubble through the liquid. The terminal velocity is $v_t = \frac{2}{9} \frac{\rho g R^2}{\mu}$~\cite{Clift}, which allows us to calculate the viscosity for different SDS concentrations. The viscosity of the three aqueous baths are 1.5 mPa$\cdot$s, 1.6 mPa$\cdot$s and 2.0 mPa$\cdot$s for SDS concentrations of 0.035 M, 0.14 M and 0.28 M. 
 
\section{Appendix B: Imaging and analysis}
The bubble chain was imaged simultaneously from two orthogonal planes, with two mirrors set at 22.5$^\circ$ to the normal plane at the back of the chamber. Fig.~\ref{f:images}(a) shows the geometry of the chamber with a micropipette inserted through the top of the bath. The micropipette produces bubbles in the center of the chamber, and the left and right mirror show the reflection. Images are taken with a lens (Edmund Optics, 2x Ultra Compact Objective) with a narrow depth of field, so that when the reflections are in-focus the physical bubbles are not visible in the center of the image. Image analysis was carried out as follows. The quantity $q$, the chain production speed, was determined by counting the number of bubbles produced for a given period of time, and multiplying by the diameter of the bubbles. The terminal speed $v$ was extracted by binarizing the image, tracking the center of the chain, then using auto-correlation between subsequent frames to extract the distance traveled in pixels. The speed was averaged over the course of the video to extract the distance moved in the vertical direction per frame. The bubble radius $R$ was found using template matching. A sample bubble is cropped, then a template matching function is used to find likely locations of the bubbles. Starting at the bottom of the frame where the bubbles are produced, we match the bubble along the chain from the right and left side and using trigonometry for the two orthogonal directions, we calculate the distance between the center of subsequent bubbles. From this distance, we can determine the radius. The amplitude of the buckling $A$ is determined by averaging all of the images for a given video. Averaging blurs the image as a function of time, and a horizontal cross section of the intensity indicates the outer edges of the bubble chain. Subtracting off the bubble radius and reconstructing the averaged data in 3-dimensions gives the average buckling amplitude. We define a buckled chain as one with a sustained buckling amplitude of greater than 0.2 $R$. We measured the wavelength with ImageJ over 5 wavelengths. Lastly, we note that at high bubble production speeds the chain can break up further away from the orifice, which we attribute to the larger viscous drag acting on the edges of the chain compared to the center. For this reason the image analysis window is limited to sections between the orifice and the location of break-up.


\begin{thebibliography}{47}
\bibitem [{tau()}]{tau}%
  \BibitemOpen
  \href@noop {} {}\bibinfo {note} {The time constant $\tau$ is proportional to
  the time it takes an isolated single bubble, traveling at its terminal
  velocity, to move through the bath a distance of 2R and sets the relevant
  timescale of the system.}\BibitemShut {Stop}%
\end{thebibliography}

%

\end{document}